\documentclass[showpacs,amsmath,amssymb,aps,prd,preprint,groupedaddress,superscriptaddress]{revtex4-1}
\usepackage[latin1]{inputenc}

\begin{document}


\title{From Brans-Dicke gravity to a geometrical scalar-tensor theory}


\author{T. S. Almeida}
\email{talmeida@fisica.ufpb.br}
\author{M. L. Pucheu}
\email{mlaurapucheu@fisica.ufpb.br}

\author{C. Romero}
\email{cromero@fisica.ufpb.br}

\affiliation{Universidade Federal da Paraíba, Departamento de Física, C. Postal 5008, 58051-970 João Pessoa, Pb, Brazil}

\author{J. B. Formiga}
\email{jansen.formiga@uespi.br}

\affiliation{Centro de Ciências da Natureza, Universidade Estadual do Piauí, C. Postal 381, 64002-150 Teresina, Piauí, Brazil}


\date{\today}

\begin{abstract}
We consider an approach to Brans-Dicke theory of gravity in which the scalar
field has a geometrical nature. By postulating the Palatini variation, we find
out that the role played by the scalar field consists in turning the
space-time geometry into a Weyl integrable manifold. This procedure leads to a
scalar-tensor theory that differs from the original Brans-Dicke theory in many
aspects and presents some new features.
\end{abstract}


\maketitle

\section{Introduction}

As is widely known, an important guide to Einstein in the development of his
general theory of relativity was what he called the principle of equivalence,
which, in mathematical terms, corresponds to the so-called geodesic postulate,
i.e., the assumption that free particles under the sole influence of gravity
will follow geodesics in a curved space-time. This, clearly, was the first
step towards a geometrization of the gravitational interaction. The second
step came from setting the field equations, which then establish how matter
curves space-time. Or, in the words of American physicist John Wheeler,
\textquotedblleft space-time tells matter how to move, and matter tells
space-time how to curve\textquotedblright. The interesting fact here is that
this elegant theoretical scheme, which has set the stage for general
relativity, works perfectly with many other metric theories of gravity,
including those whose geometrical framework is not \textit{a priori }assumed
to be Riemannian or that make use of physical variables other than the metric
and matter fields. This is the case, for instance, of one of the most popular
alternative theories of gravity, namely, Brans-Dicke scalar-tensor theory of
gravity, a theoretical framework in which the space-time manifold is still
assumed to be Riemannian, but the gravitational interaction is described by
two fields: the metric tensor $g_{\mu\nu}$ and a scalar field $\Phi$
\cite{PhysRev.124.925,*PhysRev.125.2163}. It turns out, however, that these two fields are of quite a
distinct nature. Indeed, while $g_{\mu\nu}$ is essentially geometric, $\Phi$
does not appear in the equations of motion of particles and photons. In fact,
$\Phi$ is neither a matter field nor a geometric field, and is traditionally
interpreted as the inverse of the gravitational coupling parameter, which in
Brans-Dicke theory is not constant and is considered to be determined by the
matter content of the Universe. This non-geometrical character of $\Phi$ has
led us to speculate on what kind of gravitational theory would result if
$\Phi$ were assigned an active geometrical role in the dynamics of the
gravitational field as well as in the equations of motion of particles and
light. Surely, in this case we would expect that, being part of the geometry,
$\Phi$ should appear explicitly in the geodesic equations. Moreover, in this
new scheme, the gravitational field would be described not only by $g_{\mu\nu
}$, but by the pair $(g_{\mu\nu},\Phi).$ Of course, such features would
immediately exclude Riemannian geometry as the mathematical framework to be
used to describe space-time. Instead, one would have to look into another
geometrical setting which would operate with a geometric scalar field as one
of its inbuilt fundamental constituents. This would then lead us to the
question of how to determine, from first principles, the geometry of the
space-time. Well, it seems there are at least two ways to answer this
question: one is to postulate \textit{a priori }a certain kind of geometry, as
in the case of general relativity, Brans-Dicke theory, and many others. The
second way is to chose an action and try to extract the geometry from the
action itself by means of a variational principle. As we know, there are
essentially two distinct variational principles at our disposal: the one that
uses the Hilbert method, in which the field equations are derived by
performing variations with respect to the metric, and the so-called Palatini
method, which considers independent variations of the affine connection and
the metric \cite{[{See, for instance, }] misner1973gravitation,*[{For alternative proposals to the Palatini method see }] PhysRevD.81.124019,*PhysRevD.83.101501}. It is also well known that, when applied to general
relativity, although both methods lead to the same field equations, the latter
has the additional advantage of giving a definite specification of the
Riemannian character of the space-time.  This equivalence, however, is no
longer true in the case of more general actions \cite{Faraoni_Capozziello_2011}. In view of
the this special feature, i.e., the ability of the Palatini method to
determine the space-time geometry directly from the action, it seems natural
to apply this method, and even extend it, to investigate the Brans-Dicke
action if we are to assign any geometrical role to the scalar field $\Phi$. In
the present work we begin by applying the Palatini variational method to the
Brans-Dicke action. However, because the scalar field $\Phi$ is now regarded
as an independent geometric field in its own right, we shall assume that
$\Phi$, the metric $g_{\mu\nu}$ and the connection $\Gamma_{\mu\nu}^{\alpha}$
must be varied independently. As we shall see, the field equations
corresponding to the variation of the connection will allow us to identify the
space-time geometry as a special case of Weyl geometry, with the scalar field
$\Phi$ playing the role of the Weyl field \cite{Weyl:1918ib, *weyl1952ace, *[{A nice account of Weyl's ideas as well as the refutation of his gravitational theory may be found in }] pauli1981theory, *[{See, also, }] RevModPhys.72.1, *a, *[{For a more formal mathematical treatment, see }] zbMATH03310689}. It is worth noting that
a close connection between Brans-Dicke theory and Weyl geometry has already
been discovered and may be found in different contexts. In fact, this
connection has been shown to exist for any scalar-tensor theory in which the
scalar field is non-minimally coupled to the metric \cite{[{See, for instance, }]  Novello2,doi:10.1142/S2010194511001176} .

It turns out that the change from Hilbert to (an extended) Palatini
variational principle when applied to the Brans-Dicke action will lead us to a
new scalar-tensor theory of gravity, which presents some distinct features
compared with the original Brans-Dicke gravity. For instance, it will be found
that the space-time is no longer Riemannian, but now has the geometrical
structure of what came to be known in the literature as a Weyl integrable
space-time (WIST) \cite{[{For gravitational theories formulated in WIST and related
topics, see }] Novello198310,*Bronnikov:1995ya,*doi:10.1142/S021827189200032X,*0264-9381-13-3-004,*0264-9381-14-10-010,*Melnikov542,*1475-7516-2011-11-051,*Arias2002187,*0264-9381-21-12-014,*1742-6596-8-1-017,*Israelit1725,*:/content/aip/journal/jmp/49/10/10.1063/1.3000049,*Aguilar1205,*doi:10.1142/S0217732310034201}. Moreover, the usual coupling between matter
and gravitation assumed in Brans-Dicke theory must be modified if we want the
equivalence principle to hold in the new theory. These departures from
Brans-Dicke theory lead us to a new scenario, in which the scalar field has a
geometrical meaning and plays a fundamental and active role in the motion of
particles and light.

The paper is organized as follows. In Sec. 2, we obtain the field equations
from the extended Palatini variational method, where the scalar field $\Phi$
is now reinterpreted as a purely geometric field, hence being regarded as a
fundamental component of the space-time manifold. In Sec. 3, we compare the
field equations with those of Brans-Dicke theory and show that although the
two theories are not physically equivalent they bear strong similarities. We
proceed to Sec. 4 to show that the field equations viewed in the Riemann frame
are formally equivalent to those given by the general relativistic action
corresponding to a massless scalar field minimally coupled with the
gravitation field. In Sec. 5, this correspondence between the two theories is
used to analyze some typical solar system experiments in the context of the
geometrical scalar-tensor theory. In Sec. 6, we briefly discuss the existence
of spherically-symmetric space-times by simply looking into some corresponding
general relativistic solutions and this seems to suggest that we can view
naked singularities and wormholes as geometric phenomena. We conclude with
some remarks in Sec. 7.

\section{A geometrical approach to scalar-tensor theory}

Let us start with the Brans-Dicke action \footnote{Throughout the paper we
shall use the following convention: Whenever the symbol $g$ appears in the
expression $\sqrt{-g}$ it denotes $\det g.$ Otherwise $g$ denotes the metric
tensor.}
\begin{equation}
S_{G}=\int d^{4}x\sqrt{-g}(\Phi\mathit{R}+\frac{\omega}{\Phi}\Phi^{,\alpha
}\Phi_{,\alpha}), \label{theory01}%
\end{equation}
which will be supposed to describe the gravitational field in the absence of
matter \cite{PhysRev.124.925,*PhysRev.125.2163}. Here, we are denoting $R=g^{\mu\nu}R_{\mu\nu}(\Gamma)$, and,
in what follows, we shall consider the Ricci tensor $R_{\mu\nu}(\Gamma)$ as
being entirely expressed in terms of the affine connection coefficients
$\Gamma_{\mu\nu}^{\alpha}$ through the definition of the curvature tensor
\footnote{Throughout this paper we shall adopt the following convention in the
definition of the Riemann and Ricci tensors: $R_{\;\mu\beta\nu}^{\alpha
}=\Gamma_{\beta\mu,\nu}^{\alpha}-\Gamma_{\mu\nu,\beta}^{\alpha}+\Gamma
_{\rho\nu}^{\alpha}\Gamma_{\beta\mu}^{\rho}-\Gamma_{\rho\beta}^{\alpha}%
\Gamma_{\nu\mu}^{\rho};$ $R_{\mu\nu}=R_{\;\mu\alpha\nu}^{\alpha}.$ In this
convention, we shall write the Einstein equations as $R_{\mu\nu}-\frac{1}%
{2}Rg_{\mu\nu}-\Lambda g_{\mu\nu}=-\kappa T_{\mu\nu},$ with $\kappa=\frac{8\pi
G}{c4}$.}. Changing to the new variable $\phi$ defined by $\Phi=e^{-\phi}$, it
is easily seen that (\ref{theory01}) becomes
\begin{equation}
S_{G}=\int d^{4}x\sqrt{-g}e^{-\phi}(\mathit{R}+\omega\phi^{,\alpha}%
\phi_{,\alpha}). \label{theory02}%
\end{equation}
As we have mentioned above, we want to regard the usual Brans-Dicke scalar
field $\Phi$ (or, equivalently, $e^{-\phi})$ as possessing an intrinsic
geometrical character, which, up to now, is unknown to us. We shall then apply
the extended Palatini variational method, which amounts to take independent
variations of the three geometric objects entering in the action
(\ref{theory02}), namely, $\Phi$, $\Gamma_{\mu\nu}^{\alpha}$ and $g_{\mu\nu}%
$.  Let us first take the variation of (\ref{theory02}) with respect to the
affine connection $\Gamma_{\mu\nu}^{\alpha}$. After simple calculations we obtain%

\begin{equation}
\nabla_{\alpha}(\sqrt{-g}e^{-\phi}g^{\mu\nu})=0, \label{compatibility}%
\end{equation}
which is easily verified to be equivalent to%

\begin{equation}
\nabla_{\alpha}g_{\mu\nu}=g_{\mu\nu}\phi_{,\alpha}. \label{theory06}%
\end{equation}
It turns out that the above equation expresses nothing else than the so-called
Weyl compatibility condition between the metric and the connection (also
called Weyl nonmetricity condition). In this way, we see that the scalar field
$\phi$ acquires a clear geometrical character, while the space-time is
naturally endowed with the Weyl integrable space-time. \cite{[{For gravitational theories formulated in WIST and related
topics, see }] Novello198310,*Bronnikov:1995ya,*doi:10.1142/S021827189200032X,*0264-9381-13-3-004,*0264-9381-14-10-010,*Melnikov542,*1475-7516-2011-11-051,*Arias2002187,*0264-9381-21-12-014,*1742-6596-8-1-017,*Israelit1725,*:/content/aip/journal/jmp/49/10/10.1063/1.3000049,*Aguilar1205,*doi:10.1142/S0217732310034201}.

After the determination of the space-time geometry it seems natural that the
next step is to consider a variation of the action (\ref{theory02}) with
respect to the geometric scalar field $\phi$. Strictly speaking, this amounts
to propose an extension of the Palatini variational method as now we have
three independent geometric entities, namely, the affine connection
$\Gamma_{\mu\nu}^{\alpha}$, the metric $g_{\mu\nu}$ and the scalar field
$\phi$ being involved in the process of variation. Let us briefly recall the
geometrical role played by these three fields: the metric $g_{\mu\nu}$ is
responsible for measuring lengths and angles, the connection $\Gamma$ sets the
rules for parallel transport and defines the covariant derivatives of vector
and tensor fields, whereas the scalar field $\phi$ defines the nonmetricity,
also participating in the parallel transport of vectors, modifying their
length at each point of the space-time manifold.

Before going further, some comments about the Weyl geometry are in order
\cite{Weyl:1918ib, *weyl1952ace, *[{A nice account of Weyl's ideas as well as the refutation of his gravitational theory may be found in }] pauli1981theory, *[{See, also, }] RevModPhys.72.1, *a, *[{For a more formal mathematical treatment, see }] zbMATH03310689, [{For a comprehensive review on Weyl geometry see }] Scholz:2011za, *[{See also }] Scholz:2012ev}. Broadly speaking, we can say that the geometry conceived
by Weyl is a simple generalization of Riemannian geometry. Indeed, instead of
regarding the Levi-Civita compatibility condition, Weyl has extended it to the
more general requirement
\begin{equation}
\nabla_{\alpha}g_{\mu\nu}=\sigma_{\alpha}g_{\mu\nu}, \label{compatibility1}%
\end{equation}
where $\sigma_{\alpha}$ denotes the components of a one-form field $\sigma$,
globally defined in the manifold. If $\sigma$ is an exact form, i.e.,
$\sigma=d\phi$, where $\phi$ is a scalar field, then we have what has been
called a Weyl integrable geometry. In perfect analogy with Riemannian
geometry, the condition (\ref{theory06}) is sufficient to determine the Weyl
connection $\nabla$ in terms of the metric $g$ and the Weyl scalar field.
Thus, it is not difficult to verify that the coefficients $\Gamma_{\mu\nu
}^{\alpha}$ of the affine connection when expressed in terms of $g_{\mu\nu}$
and $\phi$ are given by
\begin{equation}
\Gamma_{\mu\nu}^{\alpha}=\left\{  _{\mu\nu}^{\alpha}\right\}  -\frac{1}%
{2}g^{\alpha\beta}\left(  g_{\beta\mu}\phi_{,\nu}+g_{\beta\nu}\phi_{,\mu
}-g_{\mu\nu}\phi_{,\beta}\right)  , \label{weyl connection}%
\end{equation}
where $\left\{  _{\mu\nu}^{\alpha}\right\}  $ denotes the Christoffel symbols.

At this point, it is vitally important to note that the Weyl condition
(\ref{compatibility1}) remains unchanged when we perform the following
simultaneous transformations in $g$ and $\sigma$:%
\begin{equation}
\overline{g}=e^{f}g, \label{conformal}%
\end{equation}%
\begin{equation}
\overline{\sigma}=\sigma+df, \label{gauge1}%
\end{equation}
where $f$ is a scalar function defined on $M$. These transformations are known
in the literature as Weyl transformations. An important fact that deserves to
be mentioned is the invariance of the affine connection coefficients
$\Gamma_{\mu\nu}^{\alpha}$ under Weyl transformations, which in turn, implies
the invariance of the affine geodesics.

The set $(M,g,\phi)$ consisting of a differentiable manifold $M$ endowed with
a metric $g$ and a Weyl scalar field $\phi$ will be referred to as a
\textit{Weyl frame}. In the particular case of a Weyl integrable manifold
(\ref{gauge1}) becomes
\begin{equation}
\overline{\phi}=\phi+f. \label{gauge}%
\end{equation}
Note that if we set $f=-\phi$ in the above equation, we get $\overline{\phi
}=0$. In this case, we refer to the set $(M,\overline{g}=e^{-f}g,\overline
{\phi}=0)$ as the \textit{Riemann frame}, since in this frame the manifold
becomes Riemannian. Incidentally, it can be easily checked that
(\ref{weyl connection}) follows directly from $\nabla_{\alpha}\overline
{g}_{\mu\nu}=0$. This simple fact has interesting and useful consequences. One
consequence is that since $\overline{g}=e^{-\phi}g$ is invariant under the
Weyl transformations (\ref{conformal}) and (\ref{gauge}) any geometric
quantity constructed exclusively with $\overline{g}$ is invariant. Other
geometric objects such as the components of the curvature tensor $R_{\beta
\mu\nu}^{\alpha}$ , the components of the Ricci tensor $R_{\mu\nu}$ , the
scalar $e^{\phi}R$ are evidently invariant under the Weyl transformations
(\ref{conformal}) and (\ref{gauge}).

It is important to note here that because the Weyl transformations
(\ref{conformal}) and (\ref{gauge}) define an equivalence relation between
frames $(M,g,\phi)$ it seems more natural to focus our attention on the
equivalence class of such frames rather than on a particular one. In this
regard, a Weyl manifold may be regarded as a frame $(M,g,\phi)$ that is only
defined \textquotedblleft up to a Weyl transformation\textquotedblright. Thus
Weyl manifolds may be treated by selecting a frame in the equivalence class,
and applying only invariant constructions to the chosen frame. From this stand
point, it would be more natural to redefine some Riemannian concepts to meet
the requirements of Weyl invariance. This view point is analogous to the way
one treats \textit{conformal geometry}, a branch of geometry, in which the
geometric objects of interest are those that are invariant under conformal
transformation, such as, for instance, the angle between two directions
\cite{[{See, for instance, }] Kobayashi_1972}. In the same spirit one should naturally modify the
definition of all invariant integrals when dealing with the integration of
exterior forms. For instance, the Riemannian $p$-dimensional volume form
defined as $\Omega=\sqrt{-g}dx^{1}\wedge...\wedge dx^{p}$ , which is not
invariant under Weyl transformations, should be replaced by $\Omega=\sqrt
{-g}e^{-\frac{p}{2}\phi}dx^{1}\wedge...\wedge dx^{p}$, and so on. Accordingly,
in a Weyl integrable manifold it would be more natural to define the concept
of \textquotedblleft length of a curve\textquotedblright in an invariant way.
As a consequence, our notion of proper time as the arc length of worldlines in
four-dimensional Lorentzian space-time should be modified. In view of this, we
shall redefine the proper time $\Delta\tau$ measured by a clock moving along a
parametrized timelike curve $x^{\mu}=x^{\mu}(\lambda)$ between $x^{\mu}(a)$
and $x^{\mu}(b)$ , in such a way, that $\Delta\tau$ is the same in all frames.
This leads us to the following definition:
\begin{equation}
\Delta\tau=\int_{a}^{b}\left(  \overline{g}_{\mu\nu}\frac{dx^{\mu}}{d\lambda
}\frac{dx^{\nu}}{d\lambda}\right)  ^{\frac{1}{2}}d\lambda=\int_{a}%
^{b}e^{-\frac{\phi}{2}}\left(  g_{\mu\nu}\frac{dx^{\mu}}{d\lambda}%
\frac{dx^{\nu}}{d\lambda}\right)  ^{\frac{1}{2}}d\lambda. \label{propertime}%
\end{equation}
It should be noted that the above expression may be also obtained from the
special relativistic definition of proper time by using the prescription
$\eta_{\mu\nu}\rightarrow e^{-\phi}g_{\mu\nu}.$ Clearly, the right-hand side
of this equation is invariant under Weyl transformations and reduces to the
known expression of the proper time in general relativity in the Riemann
frame. We take $\Delta\tau$, as given above, as the extension to an arbitrary
Weyl frame of general relativistic clock hypothesis, i.e. the assumption that
$\Delta\tau$ measures the proper time measured by a clock attached to the particle.

It is not difficult to verify that the extremization of the functional
(\ref{propertime}) leads to the equations
\begin{equation}
\frac{d^{2}x^{\mu}}{d\lambda^{2}}+\left(  \left\{  _{\alpha\beta}^{\mu
}\right\}  -\frac{1}{2}g^{\mu\nu}(g_{\alpha\nu}\phi_{,\beta}+g_{\beta\nu}%
\phi_{,\alpha}-g_{\alpha\beta}\phi\,_{,\nu})\right)  \frac{dx^{\alpha}%
}{d\lambda}\frac{dx^{\beta}}{d\lambda}=0, \label{geodesics}%
\end{equation}
where $\left\{  _{\alpha\beta}^{\mu}\right\}  $ denotes the Christoffel
symbols calculated with $g_{\mu\nu}$. Let us recall that in the derivation of
the above equations the parameter $\lambda$ has been chosen such that%
\begin{equation}
e^{-\phi}g_{\alpha\beta}\frac{dx^{\alpha}}{d\lambda}\frac{dx^{\beta}}%
{d\lambda}=K=const. \label{constant}%
\end{equation}
along the curve, which, up to an affine transformation, permits the
identification of $\lambda$ with the proper time $\tau$. It turns out that
these equations are exactly those that yield the affine geodesics in a Weyl
integrable space-time, since they can be rewritten as
\begin{equation}
\frac{d^{2}x^{\mu}}{d\tau^{2}}+\Gamma_{\alpha\beta}^{\mu}\frac{dx^{\alpha}%
}{d\tau}\frac{dx^{\beta}}{d\tau}=0, \label{Weylgeodesics}%
\end{equation}
where $\Gamma_{\alpha\beta}^{\mu}=\left\{  _{\alpha\beta}^{\mu}\right\}
-\frac{1}{2}g^{\mu\nu}(g_{\alpha\nu}\phi_{,\beta}+g_{\beta\nu}\phi_{,\alpha
}-g_{\alpha\beta}\phi\,_{,\nu})$, according to (\ref{weyl connection}), may be
identified with the components of the Weyl connection. Therefore, the
extension of the geodesic postulate by requiring that the functional
(\ref{propertime}) be an extremum is equivalent to postulating that the
particle motion must follow affine geodesics defined by the Weyl connection
$\Gamma_{\alpha\beta}^{\mu}$. It will be noted that, as a consequence of the
Weyl compatibility condition (\ref{compatibility}) between the connection and
the metric, (\ref{constant}) holds automatically along any affine geodesic
determined by (\ref{Weylgeodesics}). Because both the connection components
$\Gamma_{\alpha\beta}^{\mu}$ and the proper time $\tau$ are invariant when we
switch from one Weyl frame to the other, the equations (\ref{Weylgeodesics})
are invariant under Weyl transformations.

As we know, the geodesic postulate not only makes a statement about the motion
of particles, but also determines the propagation of light rays in space-time.
Because the path of light rays are null curves, one cannot use the proper time
as a parameter to describe them. In fact, light rays are supposed to follow
null affine geodesics, which cannot be defined in terms of the functional
(\ref{propertime}), but, instead, they must be characterized by their
behaviour with respect to parallel transport. We shall extend this postulate
by simply assuming that light rays follow Weyl null affine geodesics.

We have hitherto considered the Brans-Dicke action in vacuum. However, before
we proceed with the variation with respect to $g_{\mu\nu}$ and $\phi$, it
turns out to be more convenient, as part of our reasoning, to complete
(\ref{theory01}) by adding an action $S_{m}$ to account for the matter fields.
Because we have already discovered that the space-time must be described by
two geometric fields, namely, $g_{\mu\nu}$ and $\phi$, it is reasonable to
expect both to couple with matter, preferably in a frame-independent way.
Perhaps a clue to the construction of $S_{m}$ is given by the fact, mentioned
earlier, that any geometric quantity constructed $\overline{g}=e^{-\phi}g$ is
invariant under the Weyl transformations (\ref{conformal}) and (\ref{gauge}).
It seems then that the sought-for action will be given by
\[
S_{m}=\kappa^{\ast}\int d^{4}x\sqrt{-\overline{g}}L_{m}(\overline{g}_{\mu\nu
},\Psi,\nabla_{R}^{(\overline{g})}\Psi),
\]
or, equivalently,
\begin{equation}
S_{m}=\kappa^{\ast}\int d^{4}x\sqrt{-g}e^{-2\phi}L_{m}(e^{-\phi}g_{\mu\nu
},\Psi,\nabla\Psi), \label{matter action}%
\end{equation}
where, as in Brans-Dicke theory, $\kappa^{\ast}=\frac{8\pi}{c^{4}}$, $L_{m}$
designates the matter Lagrangian, $\Psi$ stands generically for the matter
fields, $\nabla_{R}^{(\overline{g})}$ denotes the Riemannian covariant
derivative with respect to the metric $\overline{g}=e^{-\phi}g$, and $\nabla$
indicates the covariant derivative with respect to the Weyl affine
 connection \footnote{It is straightforward to verify that $\nabla
_{R}^{(\overline{g})}$ is completely equivalent to $\nabla$ since the Weyl
compatibility condition (\ref{theory06}) may be written as $\nabla_{\alpha
}(e^{-\phi}g_{\mu\nu})=0.$}. Note that $L_{m}(g,\phi,\Psi,\nabla\Psi)$ is
given by the prescription $\eta_{\mu\nu}\rightarrow e^{-\phi}g_{\mu\nu}$ and
$\partial_{\mu}\rightarrow\nabla_{\mu}$, where $\nabla_{\mu}$ denotes the
covariant derivative with respect to the Weyl affine connection. Let us recall
here that $L_{m}(g,\phi,\Psi,\nabla\Psi)\equiv L_{m}^{sr}(e^{-\phi}%
g,\Psi,\nabla\Psi)$, where $L_{m}^{sr}$ denotes the Lagrangian of the field
$\Psi$ in flat Minkowski space-time of special relativity.

With the purpose of obtaining the complete field equations through the
variation of the total action $S=S_{G}+S_{m}$, we now proceed to the
definition of the energy-momentum tensor in this new geometrical setting. From
the same arguments that led to the building up of the action
(\ref{matter action}), it seems natural to define the energy-momentum tensor
$T_{\mu\nu}(\phi,g,\Psi,\nabla\Psi)$ of the matter field $\Psi$, in an
arbitrary Weyl frame $(M,g,\phi)$, by the formula
\begin{equation}
\delta\int d^{4}x\sqrt{-g}e^{-2\phi}L_{m}(g_{\mu\nu},\phi,\Psi,\nabla
\Psi)=\int d^{4}x\sqrt{-g}e^{-2\phi}T_{\mu\nu}(\phi,g_{\mu\nu},\Psi,\nabla
\Psi)\delta(e^{\phi}g^{\mu\nu}), \label{energy-momentum}%
\end{equation}
where the variation on the left-hand side must be carried out simultaneously
with respect to both $g_{\mu\nu}$ and $\phi.$ In order to see that the above
definition makes sense, it must be clear that the left-hand side of the
equation (\ref{energy-momentum}) can always be put in the same form of the
right-hand side of the same equation. This can easily be seen from the fact
that $\delta L_{m}=\frac{\partial L_{m}}{\partial g^{\mu\nu}}\delta g^{\mu\nu
}+\frac{\partial L_{m}}{\partial\phi}\delta\phi=\frac{\partial L_{m}}%
{\partial(e^{\phi}g^{\mu\nu})}\delta(e^{\phi}g^{\mu\nu})$ and that
$\delta(\sqrt{-g}e^{-2\phi})=-\frac{1}{2}\sqrt{-g}e^{-3\phi}g_{\mu\nu}%
\delta(e^{\phi}g^{\mu\nu}).$

We are now ready to perform the variation of the complete action
\begin{equation}
S=\int d^{4}x\sqrt{-g}e^{-\phi}(\mathit{R}+\omega\phi^{,\alpha}\phi_{,\alpha
})+\kappa^{\ast}\int d^{4}x\sqrt{-g}e^{-2\phi}L_{m}(e^{-\phi}g_{\mu\nu}%
,\Psi,\nabla\Psi) \label{total action}%
\end{equation}
with respect to the metric $g_{\mu\nu}.$ A simple calculation yields
\begin{equation}
R_{\mu\nu}-\frac{1}{2}g_{\mu\nu}R=-\kappa^{\ast}T_{\mu\nu}-\omega\phi_{,\mu
}\phi_{,\nu}+\frac{\omega}{2}g_{\mu\nu}\phi^{,\alpha}\phi_{,\alpha},
\label{theory24}%
\end{equation}
where it should be kept in mind that we are denoting by $R_{\mu\nu}$ and $R$
the Ricci tensor and the scalar curvature, respectively, as defined with
respect to the Weyl connection (\ref{weyl connection}). Finally, if we now
carry out the variation of the action (\ref{total action}) with respect to the
scalar field $\phi$, we obtain
\begin{equation}
R+3\omega\phi^{,\alpha}\phi_{,\alpha}+2\omega\square\phi=\kappa^{\ast}T,
\label{theory21}%
\end{equation}
where $T=g^{\mu\nu}T_{\mu\nu}$ and $\square$ denotes the d'Alembert operator
defined with respect to the Weyl connection \footnote{In the derivation of
this equation we have made use of the identity $\square\phi=\tilde
{\square}\phi-2\phi_{,\alpha}\phi^{,\alpha}$ , in which $\tilde{\square}$
denotes the usual d'Alembert operator. It should also be noted that, since
$\phi$ and $g^{\mu\nu}$ are being regarded as independent, then, when
considering the variation with respect to $\phi$, we must have $\delta
(e^{\phi}g^{\mu\nu})=e^{\phi}g^{\mu\nu}\delta\phi.$ Thus, one can easily see
that $\delta_{\phi}S_{m}=\int d^{4}x\sqrt{-g}e^{-\phi}T_{\mu\nu}g^{\mu\nu
}\delta\phi=\int d^{4}x\sqrt{-g}e^{-\phi}T\delta\phi.$}. If we now take the
trace of (\ref{theory24}) we will get
\begin{equation}
R+\omega\phi^{,\alpha}\phi_{,\alpha}=\kappa^{\ast}T \label{theory20}%
\end{equation}
which combined with (\ref{theory21}) leads to
\begin{equation}
\square\phi+\phi^{,\alpha}\phi_{,\alpha}=0. \label{theory25}%
\end{equation}

Of course we can rewrite all the field equations derived above in a Riemannian
form. All we have to do is to express the Weylian geometric quantities
$R_{\mu\nu}$ and $R$ in terms of their Riemannian counterparts, which will be
denoted by $\widehat{R}_{\mu\nu}$ and $\widehat{R}$, both calculated directly
from the metric $g_{\mu\nu}$ and the Christoffel symbols $\left\{
_{\alpha\beta}^{\mu}\right\}  $. In this way, after some straightforward
calculations and taking into account (\ref{weyl connection}) we can rewrite
(\ref{theory24}), (\ref{theory21}) and (\ref{theory25}), respectively, as%
\begin{equation}
\widehat{R}_{\mu\nu}-\frac{1}{2}g_{\mu\nu}\widehat{R}=-\kappa^{\ast}T_{\mu\nu
}-\frac{w}{\Phi^{2}}(\Phi_{,\mu}\Phi_{,\nu}-\frac{1}{2}g_{\mu\nu}\Phi
_{,\alpha}\Phi^{,\alpha})-\frac{\Phi_{\mu;\nu}}{\Phi}, \label{bd1}%
\end{equation}%
\begin{equation}
\widehat{R}+\frac{w}{\Phi^{2}}\Phi_{,\alpha}\Phi^{,\alpha}=\kappa^{\ast}T,
\label{bd2}%
\end{equation}%
\begin{equation}
\widehat{\square}\Phi=0, \label{scalarfield}%
\end{equation}
where $w=\omega-\frac{3}{2}$, $\widehat{\square}$ denotes the d'Alembert
operator defined with respect to the Riemannian connection, and, in order to
make comparisons with the Brans-Dicke field equations, we are working with the
field variable $\Phi$ $=$ $e^{-\phi}$.

\section{Similarities with Brans-Dicke theory}

The equations (\ref{theory24}), (\ref{theory21}) and (\ref{theory25}), which
we have derived in the previous section, bear strong similarities to the field
equations of Brans-Dicke theory. In fact, connections between gravity theories
based on Weyl integrable geometry and Jordan-Brans-Dicke theories are known to
exist and have already been pointed out in the literature (see, for instance,
\cite{[{For a comprehensive review on Weyl geometry see }] Scholz:2011za, *[{See also }] Scholz:2012ev}). Let us recall that Brans-Dicke field equations may be written
in the form \cite{PhysRev.124.925,*PhysRev.125.2163}
\begin{equation}
\widehat{R}_{\mu\nu}-\frac{1}{2}g_{\mu\nu}\widehat{R}=-\frac{\kappa^{\ast}%
}{\Phi}T_{\mu\nu}-\frac{\omega}{\Phi^{2}}(\Phi_{,\mu}\Phi_{,\nu}-\frac{1}%
{2}g_{\mu\nu}\Phi_{,\alpha}\Phi^{,\alpha})-\frac{1}{\Phi}(\Phi_{,\mu;\nu
}-g_{\mu\nu}\square\Phi), \label{BD1}%
\end{equation}%
\begin{equation}
\widehat{R}-2\omega\frac{\square\Phi}{\Phi}+\frac{\omega}{\Phi^{2}}%
\Phi_{,\alpha}\Phi^{,\alpha}=0, \label{BD2}%
\end{equation}
where we are keeping the notation of the previous section, in which
$\widehat{R}_{\mu\nu}$ and $\widehat{R} $ denotes the Ricci tensor and the
curvature scalar calculated with respect to the metric $g_{\mu\nu}.$ By
combining (\ref{BD1}) and (\ref{BD2}) we can easily derive the equation
\begin{equation}
\square\Phi=\frac{\kappa^{\ast}T}{2\omega+3}, \label{BD3}%
\end{equation}
which is the most common form of the scalar field equation usually found in
the literature \cite{[{We are adopting the same convention as in }] adler1975introduction}. In this way, we see that in the vacuum case,
i.e., when $T_{\mu\nu}=0$, Brans-Dicke field equations are formally identical
to (\ref{bd1})\ and (\ref{scalarfield}) if we set $w=\omega-\frac{3}{2}$.
However, the two theories are not physically equivalent since in Brans-Dicke
theory test particles follow metric geodesics and not affine Weyl geodesics.

\section{Similarities with Einstein's gravity}

In developing a geometric scalar-field gravity theory, we have hitherto
confined ourselves to a generic Weyl frame $(M,g,\phi)$, that is, a frame in
which space-time is regarded as a differentiable manifold $M$ endowed with a
metric $g$ and a non-null Weyl scalar field $\phi$. We now wonder how the
action and, consequently, the field equations will be affected if we go to the
Riemann frame $(M,\overline{g}=e^{-\phi}g,\overline{\phi}=0)$. To carry out
the change of frames, let us apply the Weyl transformations (\ref{conformal})
and (\ref{gauge}), with $f=-\phi$ to (\ref{theory02}). It is not difficult to
verify that in the new frame the action reads \footnote{At this point the
reader may wonder why a term involving $\phi$ still remains in the action in a
frame where there is no Weyl field. As a matter of fact, after the Weyl
transformation being carried out the remaining $\phi$ no longer represents the
Weyl field, which completely vanishes in the new frame (i.e., $\overline{\phi
}=0$). The presence of the term involving $\phi$ must be regarded as a mere
trace left out in the action by the specific Weyl transformation, which
implicit involves the scalar field. Thus, in the Riemann frame $\phi$ no
longer plays a geometrical role, and accordingly might be interpreted as a
physical field.}
\begin{equation}
S=\int d^{4}x\sqrt{-\overline{g}}(\overline{R}(\overline{g},0)+w\phi^{,\alpha
}\phi_{,\alpha})+S_{m}(\overline{g},\Psi,\nabla^{\overline{g}}\Psi),
\label{theory22}%
\end{equation}
where $\overline{R}(\overline{g},0)=\overline{g}^{\mu\nu}\overline{R}_{\mu\nu
}(\overline{g},0)$ are purely Riemannian terms (as $\overline{\phi}=0$) and we
are denoting $\phi^{,\alpha}\phi_{,\alpha}=\overline{g}^{\alpha\beta}%
\phi_{,\alpha}\phi_{,\beta}$. It is clear that, by construction, the matter
action and the energy-momentum tensor $T_{\mu\nu}$ are invariant with respect
to these transformations,\ that is, $S_{m}(\overline{g},0)=S_{m}(g,\phi)$ and
$T_{\mu\nu}(\overline{g},0)=T_{\mu\nu}(g,\phi)$. On the other hand, if we
rescale the scalar field $\phi$ by defining the new field variable
$\varphi=\sqrt{w}\phi$, we are finally left with the following equations
\footnote{Here we are restricting ourselves to the class of solutions with
$w>0$.}:
\begin{equation}
\overline{R}_{\mu\nu}-\frac{1}{2}\overline{g}_{\mu\nu}\overline{R}%
=-\kappa^{\ast}T_{\mu\nu}-\varphi_{,\mu}\varphi_{,\nu}+\frac{1}{2}\overline
{g}_{\mu\nu}\varphi^{,\alpha}\varphi_{,\alpha}, \label{theory16}%
\end{equation}
and
\begin{equation}
\overline{\square}\varphi=0, \label{theory19}%
\end{equation}
where $\overline{R}_{\mu\nu}$, $\overline{R}$ and $\overline{\square}\varphi$
are all defined with respect to the metric $\overline{g}=e^{-\phi}g$.
Therefore, the field equations of this geometric scalar-field theory, viewed
in the Riemann frame, are given by the general relativistic action
corresponding to a massless scalar field minimally coupled with the
gravitation field, with the only proviso that the Einstein constant $\kappa$
must be replaced by $\kappa^{\ast}$.

\section{Spherically-symmetric solutions}

Once one has set up a theory of gravity the first question to be addressed is
whether the predictions of the new proposal are in agreement with the
so-called solar-system experiments. In the case of the present geometrical
approach to scalar-tensor theory, we have seen in the previous section that
the mathematical formalism of Weyl transformations allows us to establish a
close connection of the theory with Einstein's gravity minimally coupled with
a massless scalar field. We shall take advantage of this fact to briefly
investigate the existence of spherically-symmetric space-times by simply
looking into some corresponding general relativistic solutions already known
in the literature.

Scalar fields in general relativity have long been studied with great
interest, usually as classical approximation to some effective field theory.
Also, many attempts at unifying gravity with other interactions, from
Kaluza-Klein theories to superstrings models, predict the existence of a
massless scalar field, not to mention that, according to the standard model,
the Higgs boson is described by a scalar field \cite{[{An interesting proposal that shows how scalar fields
non-minimally coupled with gravity can act as an inflaton in the early
Universe may be found in }] 0264-9381-30-21-214001}. Historically, the
first static spherically symmetric solution of the coupled
Einstein--massless-scalar-field equations was found by Fisher \cite{Fisher:1948yn}.
This solution was later rediscovered by some authors and now it is often
referred to as the Janis-Newman-Winicour solution \cite{PhysRevD.24.839,*Roberts907,*PhysRevLett.20.878}. A
generalization of Fisher solution to $n$ dimensions ($n\geqq4$) was recently
obtained in \cite{PhysRevD.40.2564} and further analyzed in details in
\cite{PhysRevD.81.024035}.

The connection between the mathematical framework of the geometrical
scalar-tensor theory and that of general relativity sourced by a massless
scalar field leads naturally to the question of to what extent the physics
described in one framework may be transported to the other. (This point remind
us of the controversial issue regarding the equivalence between the so-called
Jordan and Einstein frames in scalar-tensor theory and in $f(R)$ cosmology
\cite{PhysRevD.75.023501,*Brown:2011eh,*0264-9381-21-15-N02}.) With regard to physical phenomena that depend solely on the
motion of particles moving under the influence of gravity alone or on the
propagation of light rays, both descriptions are completely equivalent. The
reason for this lies on the fact that geodesics are invariant with respect to
Weyl transformations, hence the causal structure of space-time remains
unchanged in all Weyl frames. Moreover, as a consequence of the
above-mentioned connection between the two frameworks all results concerning
the classical solar system tests of gravity predicted by Fisher solution may
be carried over automatically to the geometrical approach.

Thus, let us consider the static, spherically symmetric vacuum asymptotically
flat solution of the field equations (\ref{theory16}) and (\ref{theory19}). As
we have mentioned above, this solution, denoted here by $\overline{g}_{\mu\nu
}$ , was first found by Fisher and its line element may be written as:%

\begin{equation}
d\overline{s}^{2}=(1-\frac{r_{0}}{r})^{\frac{M}{\eta}}dt^{2}-(1-\frac{r_{0}%
}{r})^{-\frac{M}{\eta}}dr^{2}-r^{2}(1-\frac{r_{0}}{r})^{1-\frac{M}{\eta}%
}(d\theta^{2}+\sin^{2}\theta d\psi^{2}), \label{Fisher}%
\end{equation}

\begin{equation}
\varphi=\frac{\Sigma}{\eta\sqrt{2}}\ln|1-\frac{r_{0}}{r}|,
\label{Fisher scalar}%
\end{equation}
where $r_{0}=2\eta$, $\eta=\sqrt{M^{2}+\Sigma^{2}}$ and $M>0$ is the body's
mass in the center of this coordinates \footnote{It is easy to see that when
$\Sigma\rightarrow0$, Fisher solution goes over to the Schwarzschild
space-time.}. It turns out that by using the parametrized post-Newtonian
formalism it has been shown that for a wide range of values of the massless
scalar field $\Sigma$ the Fisher solution predicts the same effects on
solar-system experiments as the Schwarzschild solution does \cite{PhysRevD.83.087502} . We
therefore conclude that, as far as solar-system experiments are concerned, due
to invariance of the geodesics under change of frames the geometrical
scalar-tensor theory yields the same results predicted by general relativity.

\section{Naked singularities and wormholes as geometrical phenomena}

The possibility of converting the present geometrical version of scalar-tensor
theory into general relativity plus a massless scalar field brings up some
interesting points. As is well known, it has been shown that the presence of a
massless scalar field in general relativity causes the event horizons of
Schwarzschild, Reissner-Nordström and Kerr solutions to be reduced to a point,
and hence leading to the appearance of naked singularities \cite{PhysRevD.31.1280,*doi:10.1142/S0217732311035602}. In
fact, naked singularities, which were predicted to appear in the process of
spherically symmetric collapse of a massless scalar field, has later been
found in other systems, such as axisymmetric gravitational waves, radiation
and perfect fluids, and so on \cite{PhysRevLett.70.9,*PhysRevLett.70.2980,*PhysRevLett.72.1782,*0264-9381-17-4-303}.

In the case of Fisher solution, given by (\ref{Fisher}), (\ref{Fisher scalar}%
), the invariant scalar $\overline{R}(\overline{g},0)=\overline{g}^{\mu\nu
}\overline{R}_{\mu\nu}(\overline{g},0)$ gives \cite{PhysRevD.81.024035}
\begin{equation}
\overline{R}=\frac{\Sigma^{2}}{r^{4}}(1-\frac{r_{0}}{r})^{(\frac{M}{\eta}-2)},
\label{invariant}%
\end{equation}
which means we have a naked singularity at $r=r_{0}$, since $\frac{M}{\eta
}=\frac{M}{\sqrt{M^{2}+\Sigma^{2}}}<1.$

It is important to note that the scalar (\ref{invariant}), obtained from
(\ref{Fisher}), may be looked upon as the Weyl invariant $e^{\phi}R$ ,
calculated in the Riemann frame $(M,\overline{g},$ $\overline{\phi}=0)$. As we
have already pointed out in Section 2, this scalar is invariant under the Weyl
transformations (\ref{conformal}) and (\ref{gauge}). This means that if we go
back to the Weyl frame $(M,g,\phi)$, where the field equations are
(\ref{theory20}) and (\ref{theory25}), we still have a space-time singularity
at $r=r_{0}$.

It is interesting to have a look at Fisher solution when viewed in the Weyl
frame $(M,g,\phi).$ The Weyl transformation that does this task leads to the
metric given by
\[
g_{\mu\nu}=e^{\phi}\overline{g}_{\mu\nu}=e^{\frac{\varphi}{\sqrt{w}}}%
\overline{g}_{\mu\nu}=e^{\frac{\Sigma}{\eta\sqrt{2w}}\ln|1-\frac{r_{0}}{r}%
|}\overline{g}_{\mu\nu}\text{ ,}%
\]
whereas $\phi=\frac{\Sigma}{\eta\sqrt{2w}}\ln|1-\frac{r_{0}}{r}|$ is the
geometric scalar field in this frame. The line element corresponding to this
metric will be written as
\begin{equation}
ds^{2}=(1-\frac{r_{0}}{r})^{\frac{M}{\eta}+\frac{\Sigma}{\eta\sqrt{2w}}}%
dt^{2}-(1-\frac{r_{0}}{r})^{-\frac{M}{\eta}+\frac{\Sigma}{\eta\sqrt{2w}}%
}dr^{2}-r^{2}W^{(1-\frac{r_{0}}{r})1-\frac{M}{\eta}+\frac{\Sigma}{\eta
\sqrt{2w}}}(d\theta^{2}+\sin^{2}\theta d\psi^{2}). \label{fisher-weyl}%
\end{equation}
In order to see that we still have a naked singularity in the frame
$(M,g,\phi)$ let us recall that the area of the surface $\Gamma$ defined by
$t=const,$ $r=r_{0}$ must be calculated with the invariant integral $A=%
{\displaystyle\int\limits_{\Gamma}}
e^{-\phi}\sqrt{\left\vert h\right\vert }d\theta\wedge d\psi$, where $h$
denotes the determinant of metric on $\Gamma$ induced by (\ref{fisher-weyl}).
Since $A$ is invariant under Weyl transformations and $A=0$ in the Riemann
frame we conclude that (\ref{fisher-weyl}) indeed represents a space-time with
a naked singularity.

It is well known that the existence of naked singularities in Fisher
space-time is a consequence of the presence of a massless scalar field, a
field that is related to a massless particle of zero spin. Up to now no such
particles have been discovered and all known spin zero particles are massive,
hence models with massless scalar fields do not seem to be realistic. Also, it
is still not clear whether such a solution can be considered as a result of
gravitational collapse, thereby representing a violation of the cosmic
censorship conjecture \cite{penrcoscen,*Israel1049}. In the present approach, however, it
should be noted that the scalar field is not a physical field, but should be
regarded as an essential part of the geometric structure of space-time.
Violation of the cosmic censorship conjecture in this case occurs in quite a
different context compared with its general relativistic counterpart.

It should be noted that in deriving the field equations (\ref{theory16}) we
have implicitly considered $w>0$. If we do not want to impose any restriction
on the value of $w$ it is preferable to work with the field variable $\phi$,
and in this case the field equations in the Riemann frame reads
\begin{equation}
\overline{R}_{\mu\nu}-\frac{1}{2}\overline{g}_{\mu\nu}\overline{R}%
=-\kappa^{\ast}T_{\mu\nu}-w\phi_{,\mu}\phi_{,\nu}+\frac{w}{2}\overline{g}%
_{\mu\nu}\phi^{,\alpha}\phi_{,\alpha}, \label{massless scalar 1}%
\end{equation}

\begin{equation}
\overline{\square}\phi=0. \label{massless scalar 2}%
\end{equation}
These equations, in which the coupling constant $w$ appears explicitly, were
first considered by Bergmann and Leipnik \cite{PhysRev.107.1157}. The most general
spherically symmetric solution to the coupled Einstein-massless-scalar-field
equations were obtained by Wyman and, in fact, includes Fisher's solution as a
particular case \cite{PhysRevD.24.839,*Roberts907,*PhysRevLett.20.878}. The line element and the scalar field
corresponding to Wyman's solution are given by
\begin{equation}
d\overline{s}^{2}=e^{\frac{\alpha}{R}}dt^{2}-e^{-\frac{\alpha}{R}}\left[
\frac{\frac{\gamma}{R}}{\sinh(\frac{\gamma}{R})}\right]  ^{4}dR^{2}%
-R^{2}e^{-\frac{\alpha}{R}}\left[  \frac{\frac{\gamma}{R}}{\sinh(\frac{\gamma
}{R})}\right]  ^{2}(d\theta^{2}+\sin^{2}\theta d\psi^{2}), \label{Wyman1}%
\end{equation}%
\begin{equation}
\phi=\frac{1}{R}, \label{Wyman2}%
\end{equation}
where $\alpha$ and $\gamma$ are constants and $\gamma=\frac{(\sqrt{\alpha
^{2}+2w})}{2}$. It is not difficult to verify that Fisher solution is a
particular case of Wyman solution for $w>0$. Indeed, if we define the
coordinate $R$ by $e^{-\frac{2\eta}{R}}=1-2\eta/r$ it is straightforward to
see that (\ref{Fisher}) and (\ref{Fisher scalar}) become, respectively,
\begin{equation}
d\overline{s}^{2}=e^{-\frac{2M}{R}}dt^{2}-e^{\frac{2M}{R}}\left[  \frac
{\frac{\eta}{R}}{\sinh(\frac{\eta}{R})}\right]  ^{4}dR^{2}-R^{2}e^{\frac
{2M}{R}}\left[  \frac{\frac{\eta}{R}}{\sinh(\frac{\eta}{R})}\right]
^{2}(d\theta^{2}+\sin^{2}\theta d\psi^{2})\text{,} \label{FisherW1}%
\end{equation}%
\begin{equation}
\phi=-\Sigma\sqrt{\frac{2}{w}}\frac{1}{R}\text{,} \label{FisherW2}%
\end{equation}
recalling that $\varphi=\sqrt{w}\phi$. Therefore, if we set $\alpha=-2M$ ,
$\gamma=\eta$, we see that Fisher's solution reduces to Wyman's solution
provided that $w>0$ and $\Sigma=-\sqrt{\frac{w}{2}}$. Incidentally, it has
been shown that Wyman's solution leads to three types of space-times according
to the value assigned to $w$: a naked singularity $(w>0)$, a Schwarzschild
black hole $(w=0)$, and a wormhole solution $(-2M^{2}$ $<w<0)$ \cite{:/content/aip/journal/jmp/46/6/10.1063/1.1920308}%
. It should be clear that all these configurations carry over to the Weyl
frame $(M,g,\phi)$.

\section{Summary and discussion}

The fact that in Brans-Dicke theory of gravity the scalar field has no
geometrical origin, while in general relativity the gravitational sector of
the action is purely geometric, has motivated some authors to look for what we
might call a geometric scalar-tensor theory of gravitation. Let us briefly
comment on some of the attempts that are known to us. In one of them, the
scalar field is given a geometrical interpretation in the spirit of the
Rainich-Misner-Dicke geometrization of the electromagnetism, although it is
restricted to the vacuum case \cite{:/content/aip/journal/jmp/10/6/10.1063/1.1664930}. In another approach, it is shown
that Brans-Dicke scalar field can be derived from pure geometry if the
space-time geometry is assumed to be the Lyra manifold \cite{Lyra52,*0264-9381-5-11-011, *[{See also }] Sen311}. Weyl
integrable geometry also appears in a scalar-tensor theory which is directly
obtained from general relativity by writing the gravitational sector of
Einstein-Hilbert action in an arbitrary Weyl frame \cite{Ross157}. (It can be
shown, however, that the resulting theory, which does not consider matter
couplings, is completely equivalent to general relativity, and is also
conformally related to Brans-Dicke theory for $\omega=-\frac{3}{2}$, hence
implying that the scalar field has no dynamics \cite{0264-9381-29-15-155015,*[{See also }]   ANDP:ANDP200610230}.) Finally, a
geometrical scalar-tensor theory was constructed using a non-Riemannian
geometry, in which the scalar field is related to a scalar torsion field
\cite{:/content/aip/journal/jmp/15/12/10.1063/1.1666603} \footnote{Let us note here that it has been shown recently that
scalar torsion may also play a role in recasting general relativity as a
scalar-tensor theory in an arbitrary \textquotedblleft Cartan
gauge\textquotedblright defined by the group of the so-called Einstein's
$\lambda$-transformation \cite{Fonseca1579}.}. In this case, the theory does not
consider the matter coupling and the vacuum field equations are identical to
those of Brans-Dicke theory written in the Einstein frame.

In this work we have also developed a scalar-tensor theory in which the scalar
field plays a definite geometrical role in the description of the
gravitational field. Basically, our procedure consists in considering the
original formulation of Brans-Dicke theory as a starting point and modifying
it by letting the Palatini formalism decide what kind of geometry should we
assign to space-time. This leads in quite a natural way to Weyl geometry,
which possesses an interesting property, namely, the invariance of geodesics
under a well defined group of transformations. In fact, this suggests that we
are concerned here with a whole class of geometries, or space-time manifolds,
that are related by a Weyl transformation. According to this view, it seems
natural that the geometric objects of interest are those that are invariant
under the invariance group of transformations. By following consistently this
idea we are naturally led to redefine our familiar notions of proper time,
space-time singularities, etc, in a way that these notions retain their
invariance character, i.e., they must be the same in all frames. Surely, this
approach will lead to new physical insights as far as a gravitational theory
constructed in this framework is concerned. Consider, for instance, the
principle of equivalence. It is clear that it will hold in every Weyl frame
inasmuch as geodesics do not change by a Weyl transformation. Of course we
have quite a distinct situation in the original formulation of Brans-Dicke
theory of gravity as regards to change of frames \footnote{Let us recall here
that the mathematical transformations that relate Brans-Dicke solutions in the
Jordan and Einstein frames, defined only for $\omega>0$, are very similar to
Weyl transformations although restricted to Riemannian space-times.}. For
example, it is widely known that freely falling particles do not move on
geodesics in the so-called Einstein (conformal) frame and also measurements
made by rods and clocks are not invariant under a change of frames
\cite{Faraoni2004}.

Another comment is in order. It is important to call attention for the fact
that in obtaining Eq. (\ref{compatibility}) by applying the Palatini formalism
we have completely ignored the matter action and considered only the action
corresponding to the gravitational sector.There is, in fact, a methodological
reason to justify this procedure: It is assumed, as a principle, that what
really determines the space-time geometry is the gravitational sector. Once
the geometry is found, then completing the action by later adding the matter
action will not affect (\ref{compatibility}), since any dependence on the
affine connection may be entirely reduced to dependence on the geometric
fields $g$ and $\phi$ through (\ref{weyl connection}). This permits us to
proceed with our reasoning without having to make the usual assumption that
the matter sector is functionally independent of the (non-metric) connection
\cite{Faraoni_Capozziello_2011}.

Finally, is interesting to note that the reason why the field equations
(\ref{bd1}) and (\ref{scalarfield}) derived in Section 2 coincide with those
of Brans-Dicke theory only in the case of vacuum is that in the latter the
scalar field does not participate directly in the way how matter couples with
the gravitational field. Indeed, in Brans-Dicke theory the action describing
ordinary matter is postulated to be of the form $S_{m}=\kappa^{\ast}\int
d^{4}x\sqrt{-g}L_{m}(g_{\mu\nu},\Psi,\nabla\Psi)$, which is a necessary
requirement to ensure that freely falling particles follow Riemannian
geodesics. However, in the geometrical scalar-tensor theory we are considering
freely falling particles should follow affine geodesics in a Weyl integrable
space-time and the only matter coupling which is consistent with this
requirement is the one given by (\ref{matter action}). This can easily be
seen, for instance, by considering the field equations (\ref{theory24}) in the
case where $T_{\mu\nu}$ represents the energy-momentum tensor of a
pressureless perfect fluid (\textquotedblleft dust\textquotedblright). Then,
it is not difficult to verify that by taking the covariant divergence (with
respect to the Weyl connection) of both sides of (\ref{theory24}) we are led
to (\ref{Weylgeodesics}) \footnote{It is a well known fact that the geodesic
equations may be derived directly from the field equations. This is a known
result, which goes back to Einstein and Papapetrou \cite{[{See, for instance, }] Einstein_Infeld_Hoffmann_1938,*Papapetrou23101951,*Corinaldesi23101951}. The
argument goes like this: Consider an assembly of free particles (i.e., not
interacting with each other). If there are many of them, we can consider them
together as a pressureless perfect fluid (\textquotedblleft
dust\textquotedblright). Then, a straightforward calculation shows that
(\ref{theory24}) leads to the affine geodesic equations with the connection
coefficients given by (\ref{weyl connection}) (of course, in the case of
general relativity the scalar field $\varphi$ does not appear in the
equations). Therefore, since the worldlines of the fluid particles are
geodesics and because they are not interacting with each other we can infer
that the worldline of a single free \textquotedblleft test\textquotedblright
particle is a geodesic.}.

To conclude, we would like to remark that scalar-tensor theories have been
extensively discussed in the literature. One of the most important area of
their application is cosmology, where the scalar field is sometimes considered
as a quintessence field \cite{PhysRevD.61.064007,*PhysRevD.61.043506,*PhysRevD.66.043522,*PhysRevD.63.124006}. Scalar-tensor theories have also
been investigated in the context of braneworld scenarios \cite{0034-4885-67-12-R02,*PhysRevLett.84.2778}%
. Thus, a natural  follow up of the ideas we have discussed in the present
article would be an application of the geometric scalar-tensor theory to
modern cosmology. We leave this for future work.

\section*{Acknowledgments}

The authors thanks CNPq and CLAF for financial support.


%

\end{document}